\documentclass[11pt,a4paper]{article}
\usepackage[T1]{fontenc}
\usepackage[utf8]{inputenc}
\usepackage{authblk}
\usepackage[square,numbers]{natbib}
\usepackage[margin=1.0in]{geometry}

\usepackage{fancyhdr}

\fancypagestyle{plain}{%
  \fancyhf{}%
  \fancyfoot[C]{Copyright © 2021 for this paper by its authors. Use permitted under Creative Commons License Attribution 4.0 International (CC BY 4.0).\\\textit{Proceedings of the Conference on Applied Machine Learning for Information Security}, 2021}%
}

\usepackage{lipsum}
\usepackage{todonotes}
\usepackage{multirow}
\usepackage{tabls}
\usepackage{wrapfig}
\usepackage{subfigure}
\usepackage{amsmath,amsthm,amssymb}
\usepackage{etoolbox}
\usepackage{array}
\usepackage{tikz}
\usepackage{pgfplots}
\usetikzlibrary{shapes,arrows,arrows.meta, positioning,shapes.misc,decorations.markings,pgfplots.groupplots}
\usepackage{adjustbox}
\usepackage{hyperref}
\usepackage{booktabs}
\usepackage{multicol}
\usepackage{algorithm}
\usepackage{algorithmicx}
\usepackage[noend]{algpseudocode}
\usepackage[title]{appendix}
\usepackage[font=small,skip=4pt]{caption}
\usepackage{listings}
\usepackage{graphicx}
\usepackage{url}
\usepackage{breakurl}

\DeclareMathOperator{\triu}{triu}

\DeclareMathOperator{\agreement}{Agreement}
\DeclareMathOperator{\synchronicity}{Synchronicity}
\algdef{SE}[DOWHILE]{Do}{doWhile}{\algorithmicdo}[1]{\algorithmicwhile\ #1}%
\definecolor{background}{HTML}{FFFFFF}
\lstdefinelanguage{json}{
    basicstyle=\normalfont\ttfamily,
    numbers=left,
    numberstyle=\scriptsize,
    stepnumber=1,
    showstringspaces=false,
    breaklines=false,
    frame=lines,
    backgroundcolor=\color{background}
}

\everypar{\looseness=-1} %Please try and avoid dangling words on their own lines! 
\title{Rank-1 Similarity Matrix Decomposition For Modeling Changes in Antivirus Consensus Through Time}
\author[1,2]{Robert J. Joyce\thanks{joyce\_robert2@bah.com}}
\author[1,2]{Edward Raff\thanks{raff\_edward@bah.com}}
\author[2]{Charles Nicholas\thanks{nicholas@umbc.edu}}
\affil[1]{Booz Allen Hamilton}
\affil[2]{University of Maryland, Baltimore County}
\date{} 

\begin{document}

\maketitle

\begin{abstract}
Although groups of strongly correlated antivirus engines are known to exist, 
at present there is limited understanding of how or why these correlations came to be. Using a corpus of 25 million VirusTotal reports representing over a decade of antivirus scan data, we challenge prevailing wisdom that these correlations primarily originate from "first-order" interactions such as antivirus vendors copying the labels of leading vendors. We introduce the Temporal Rank-1 Similarity Matrix decomposition (R1SM-T) in order to investigate the origins of these correlations and to model how consensus amongst antivirus engines changes over time. We reveal that first-order interactions do not explain as much behavior in antivirus correlation as previously thought, and that the relationships between antivirus engines are highly volatile. We make recommendations on items in need of future study and consideration based on our findings.
\end{abstract}

\section{Introduction}
Our work is motivated by two chronic problems in the study of malware, namely malware detection (deciding whether a file is benign or malicious) and malware family classification (determining which of many existing families a malware sample might belong to). These tasks both require labeled data, but new malware samples number in the millions each month \citep{spafford2014} and obtaining ground truth labels via manual analysis can take hours of effort per sample~\citep{10.1145/3290607.3313040}. For this reason, the vast majority of works use the aggregated results from a collection of antivirus engines as a source of scalable labeling \cite{zhu}. For example, a common approach to malware detection is antivirus thresholding, in which some minimum number of antivirus engines in a collection must detect a file as malicious in order for it to be considered malware \citep{gashi, jiang2020}. Likewise, plurality and majority voting amongst antivirus engines are popular strategies for performing malware family classification \citep{avclass, bayer}. A significant issue with these aggregation approaches is that all antivirus engines are treated as independent voters, yet prior work shows that some groups of antivirus engines make highly correlated labeling decisions \cite{kantchelian, zhu, avclass}. As is well attested within the ML literature, the use of highly correlated models provides little benefit \citep{Jacobs1991,Wolpert1992,Breiman1996}. The presence of strong correlations between some antivirus engines likely results in degraded accuracy when these voting methods are used.

Although the existence of correlations between antivirus engines is well-documented, there has been minimal study of why they exist. Present explanations include different engines created by the same company, products ``copying'' the results of leading vendors, and vendors sub-licensing their technology to others \citep{avclass, av-meter}. All of the above explanations can be considered ``first-order'' interactions, since they create a direct link between the labeling decisions of two antivirus engines. To our knowledge, no existing work has empirically confirmed whether first-order interactions are the sole cause of the correlations between antivirus engines, or whether more complex, unknown factors are also (at least in part) responsible.

An additional consideration overlooked by prior work is the volatile and adversarial nature of the malware ecosystem. Malware authors are constantly attempting to evade detection while antivirus engines are continually forced to develop new detection methods \citep{moser}. We hypothesize that the groups of antivirus engines which are highly correlated may themselves change as a function of time. However, we are aware of no prior work which has studied this possibility \cite{Raff2020a}.

Our work does not attempt to explain how correlations between antivirus engines came to exist, but instead seeks to answer questions about the nature of these correlations and how they change over time. In Section \ref{sec:related-work} we discuss the current state of research on antivirus engine dynamics. In Section \ref{sec:av-through-time} we explore consensus amongst antivirus engines and how it has changed over the course of a decade. In Section \ref{sec:r1sm-t} we introduce the Rank-1 Similarity Matrix decomposition (R1SM), which reveals first-order interactions between the constituents of a similarity matrix. The section also discusses an extension to R1SM that uses a neural network over positional embeddings to concurrently decompose a time-series of similarity matrices, which we term R1SM-T. In Section \ref{sec:av} we apply R1SM and R1SM-T to over 25 million antivirus scan reports spanning a decade in order to identify first-order interactions between the constituent antivirus engines. Our results indicate that relationships between antivirus engines are more mercurial than previously thought. Finally, in Section \ref{sec:conclusion} we discuss the impacts of our findings and conclude that future antivirus aggregation strategies should consider approaches similar to a weighted ensemble, where the weights of each antivirus engine are a function of time.

\section{Related Work}
\label{sec:related-work}

\citet{av-meter} is the earliest work we are aware of which systematically evaluates the performance of antivirus engines. The authors observed that the detection results of many antivirus engines follow those of a leading product and hypothesize that this correlation is due to copying or sharing of information. \citet{hurier} introduced several metrics for quantifying the level of consensus between a set of antivirus engines. In Section \ref{sec:synchronicity} we explore how one of these metrics, synchronicity, changes over a ten year period. \citet{kantchelian} observed that antivirus labels take time to stabilize and that vendors may change their detections to correct errors, especially false negatives. In a study of 734,000 executables first seen on VirusTotal (an online malware analysis service that scans files with a collection of antivirus engines) between Jan. 2012 and Jun. 2014, the authors measured correlation amongst the detections of a group of approximately 80 antivirus engines. They found that although some groups are highly correlated, antivirus engines lack an overall consensus. \citet{signatureminer} surveyed a dataset of 82,866 suspicious Android applications and showed that some antivirus engines also make correlated decisions when labeling malware as a particular category or family. The closest work to ours is \citet{zhu}, who re-scanned a collection of 14,000 malware samples daily for over a year in order to investigate the dynamics of antivirus detection changes. By observing which antivirus engines changed their detections with similar timing, the authors identified five groups of highly correlated antivirus engines. Furthermore, \citet{zhu} used influence modeling to identify antivirus engines which actively change their detections to match other vendors. They determined that label copying is a widespread practice in the antivirus industry. All of these works generally lead to first-order conclusions about correlation, but do not study the correlations on the same quantity of data (25 million scan reports) or length of time (ten years) that we consider in this study. 

\section{Studying Changes in Antivirus Engine Consensus Through Time}
\label{sec:av-through-time}

\begin{figure}[h]
\centering
\includegraphics[width=0.8\columnwidth,keepaspectratio]{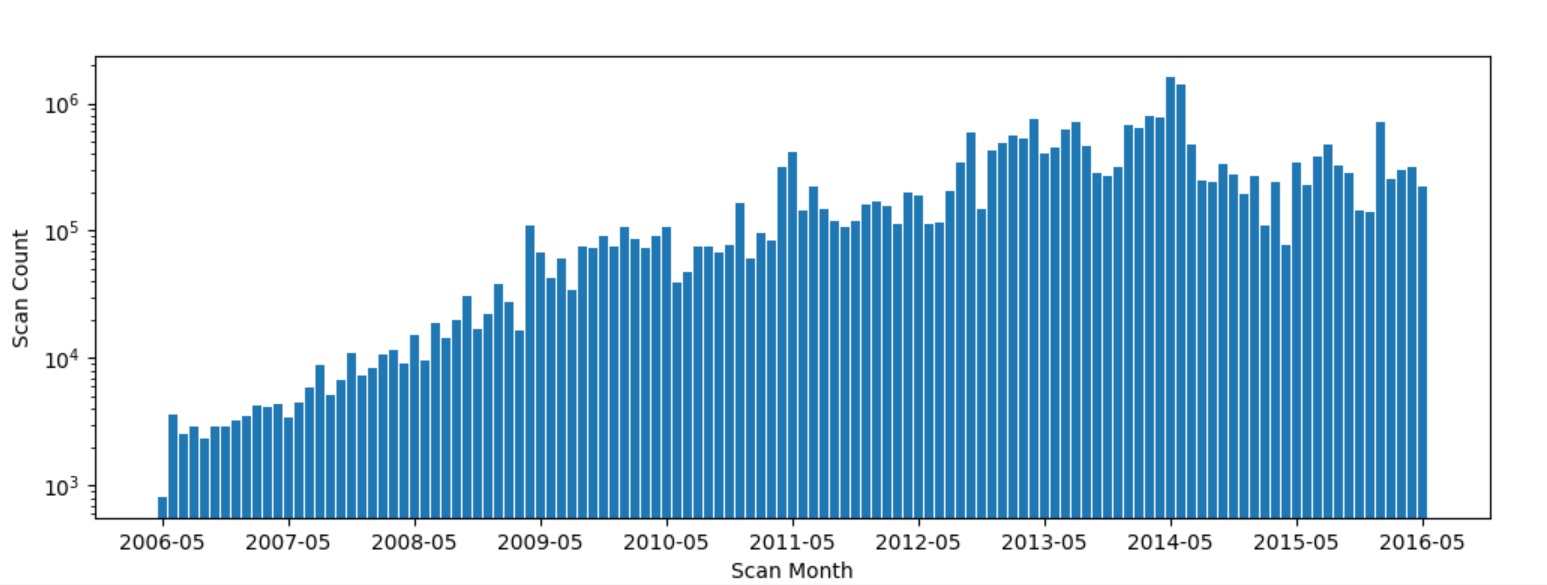}
\caption{%Histogram of the reader can see that it's a histogram
Distribution of scan dates in VirusShare-VT.}
\label{fig:datehist}
\end{figure}

For the purposes of investigating antivirus label consensus and how antivirus dynamics have changed through time, we were provided with a dataset of 25,100,286 VirusTotal scan reports \citep{seymour}. This dataset, which we call VirusShare-VT, was collected by querying the VirusTotal API for all files in chunks 0 through 233 of the publicly-available VirusShare malware corpus \citep{virusshare}. VirusTotal API queries for the VirusShare-VT dataset were made over the course of six months, from Dec. 2015 to May 2016 \citep{seymour}. Each report in the VirusShare-VT dataset is a JSON object containing information about a particular VirusTotal scan. %Figure \ref{fig:scanreport} shows the fields contained within a sample scan report. 
Of note is the \emph{scan\_date} field, which contains the date and time that a file was scanned with the collection of antivirus engines. The scan date is often older than the query date, because VT does not re-scan files for simple queries. The distribution of scan dates is shown in \autoref{fig:datehist}, ranging from May 2006 to May 2016.

Given the sizeable number of malware samples in chunks 0 - 233 of VirusShare, scanning these samples daily as \citet{zhu} did was infeasible. The VirusShare-VT dataset only contains one scan report per sample, and antivirus detections for files first seen shortly before the scan date have likely not stabilized. However, we do not consider these factors to be drawbacks, as they would be typical of most datasets used for antivirus aggregation. The massive size and timescale of the VirusShare-VT dataset makes it ideal for answering our research questions.

\subsection{Measuring Pairwise Antivirus Consensus}
\label{sec:av-consensus}

Throughout this paper we attempt to follow the terminology introduced by \citet{hurier} for measuring consensus amongst antivirus engines. Given a set of $n$ antivirus engines $A = \{a_{1}, a_{2}, ..., a_{n}\}$ and a set of $m$ files $P = \{p_{1}, p_{2}, ..., p_{m}\}$, the detections and family classifications of the antivirus engines for this set of malware samples can be arranged into two matrices $B$ and $C$:

\[
B = \begin{pmatrix} 
    b_{1,1} & b_{1,2} & \dots & b_{1,n}\\
    b_{2,1} & b_{2,2} & \dots & b_{2,n}\\
    \vdots & \vdots & \ddots & \vdots\\
    b_{m,1} & b_{m,2} & \dots & b_{m,n} 
\end{pmatrix}
\; \; \;    \; \; \; \; \; \; \; \; \;
C = \begin{pmatrix} 
    c_{1,1} & c_{1,2} & \dots & c_{1,n}\\
    c_{2,1} & c_{2,2} & \dots & c_{2,n}\\
    \vdots & \vdots & \ddots & \vdots\\
    c_{m,1} & c_{m,2} & \dots & c_{m,n} 
\end{pmatrix}
\]
\vspace*{0.1cm}

An element $B_{i,j}$ in $B$ is 1 if file $p_{i}$ is detected as malware by engine $a_{j}$ and 0 if it is not detected. An element $C_{i,j}$ in $C$ is given by the malware family assigned to file $p_{i}$ by engine $a_{j}$. $D_{i,j}$ and $C_{i,j}$ are $\emptyset$ (null) if engine $a_{j}$ did not scan $p_{i}$. For constructing the matrix $C$ we employed a portion of the AVClass labeler's architecture, which can extract family information from antivirus signatures \cite{avclass}. When AVClass ingests a scan report, it normalizes and tokenizes each antivirus signature, removes any tokens that do not contain family information, and performs family alias resolution. The processed token(s) from the antivirus signature produced by engine $a_{j}$ for file $p_{i}$ are used as the family for element $C_{i,j}$.

\citet{hurier} proposed a metric called {\em overlap} for computing pairwise detection consensus for a pair of antivirus engines. However, overlap does not consider that some antivirus engines may be missing from a scan report. Instead, we define a similar metric, which we call {\em agreement}, that corrects this issue.

\newtheorem{definition}{Definition}
\theoremstyle{definition}
\begin{definition} \label{def:agreement}
$\agreement(B_{i},B_{j}) = \frac{\left|B_{i} \; == \; B_{j}\right|}{\left|B_{i} \bigcup B_{j}\right|}\; \operatorname{s.t.}\; B_{i},B_{j} \neq \emptyset$
\end{definition}

$\agreement(B_{i},B_{j})$ divides the number of scans in $B$ in which $a_{i}$ and $a_{j}$ agree upon a file's detection by the total number of scans in which both $a_{i}$ and $a_{j}$ are present. Classification agreement can be defined in the same way by substituting the matrix $B$ for $C$.
Since it is possible for AVClass to convert a single antivirus signature into multiple family tokens, we consider two elements in $C$ to be equal if they share any AVClass tokens, or if AVClass produced zero tokens for both signatures.

\subsection{Antivirus Agreement in VirusShare-VT}
The VirusShare-VT dataset contains 93 antivirus engines that appear in at least 1,000 different scan reports. The set of antivirus engines used by VirusTotal changes gradually over time. In May 2006 only 26 of the 93 engines were observed; this number gradually increases to 57 by May 2016. The sets of antivirus engines in VirusTotal are relatively consistent month-to-month, with an average of 1.033 engines added or removed per month. Several antivirus engines only appear in VirusShare-VT during a short window of time. Many of these are alternative or beta versions of existing engines (\emph{e.g.} PandaBeta from Feb. 2007 to Feb. 2009, McAfee+Artemis from Nov. 2008 to Jan. 2011, and Avast5 from Mar. 2010 to Sep. 2011). The name,  numeric index used in all appropriate figures, and total number of occurrences of each of the 93 antivirus engines in the VirusShare-VT dataset is shown in Table \ref{tab:av-index} in Appendix A. 

\begin{figure}[ht]
    \centering
    \subfigure[Detection Matrix]{\label{fig:heatmap_detect}
    \includegraphics[width=0.40\columnwidth,keepaspectratio]{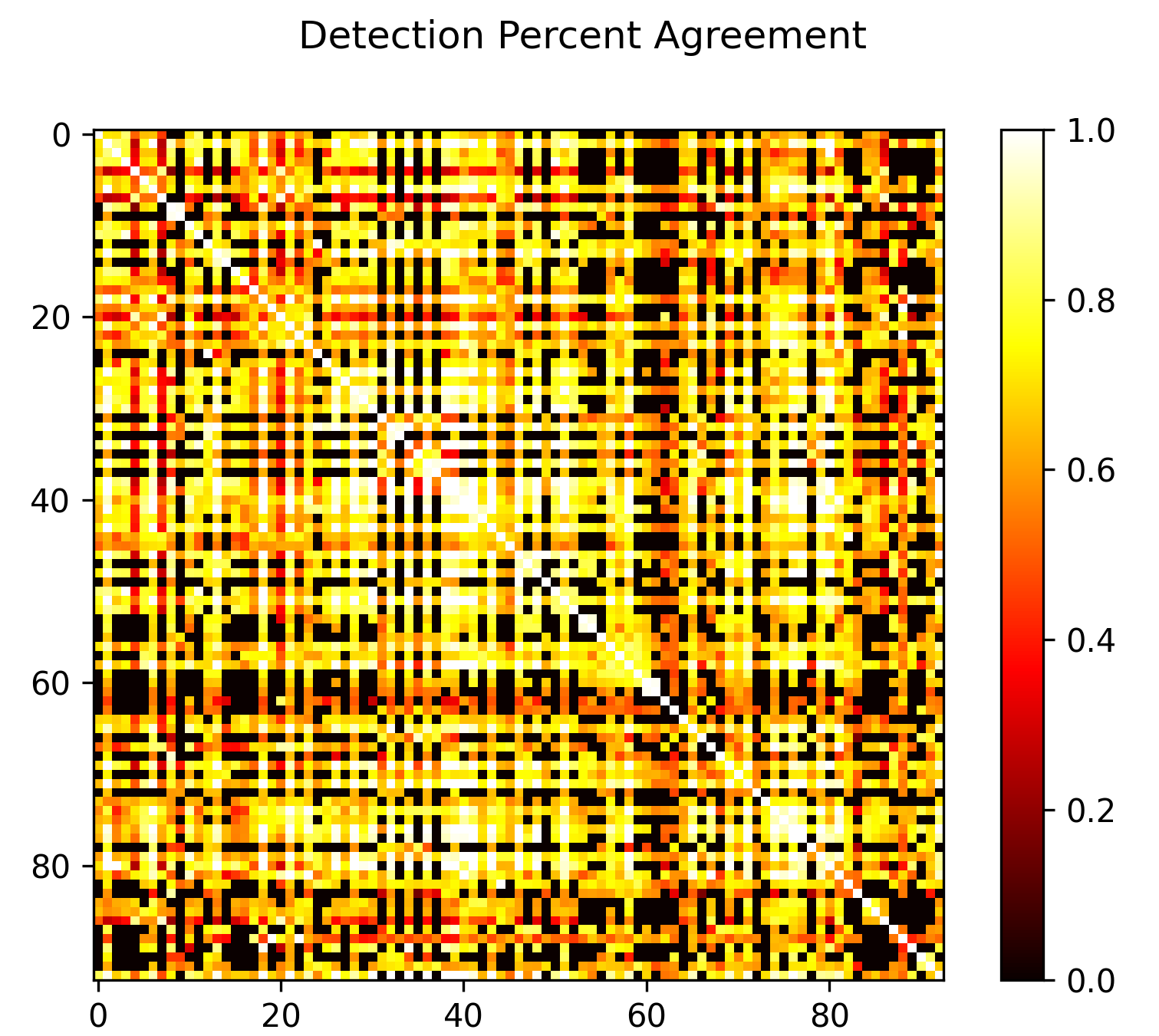}
    }
    \subfigure[Classification Matrix]{\label{fig:heatmap_family}
    \includegraphics[width=0.40\columnwidth,keepaspectratio]{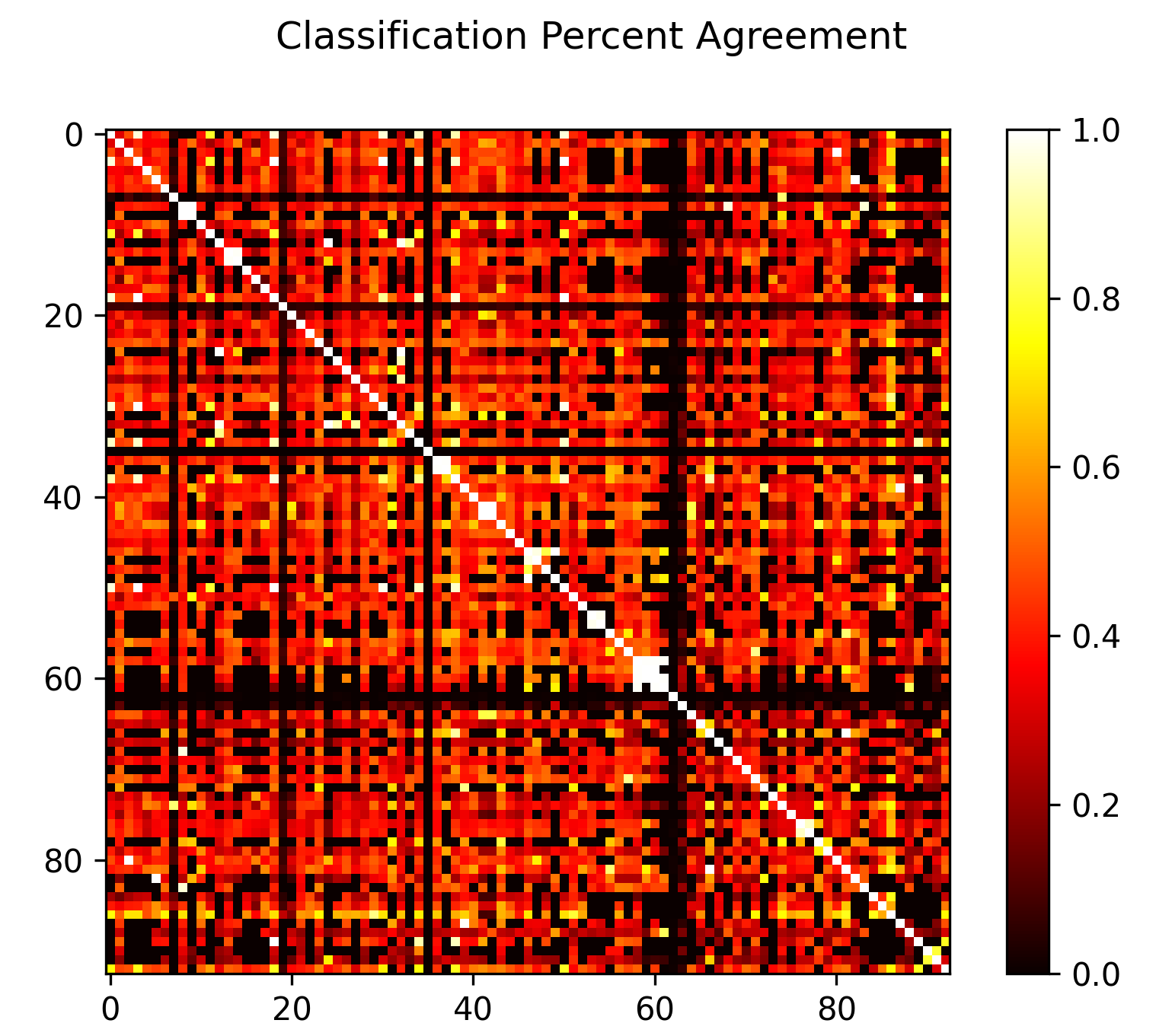}
    }
    \caption{Similarity matrices displaying pairwise detection and classification agreement for 93 antivirus engines in VirusShare-VT.}
\end{figure}

Figures \ref{fig:heatmap_detect} and \ref{fig:heatmap_family} show the pairwise detection and classification agreement for each of these 93 antivirus engines. Consistent with prior work, there are observable instances of high detection consensus among some vendors, and a small subset of vendors have very little agreement with  others \cite{kantchelian}. The classification agreement matrix appears highly similar in structure to the detection matrix but with smaller values on average. One possible explanation for this phenomenon is that classification agreement depends upon both antivirus engines detecting the sample as malware.

\subsection{Measuring Changes in Antivirus Synchronicity Over Time}
\label{sec:synchronicity}

Next, we explore how overall consensus amongst antivirus engines has changed over time. Consider a similarity matrix $D$ constructed by applying some similarity function $\textit{sim}(B_{i}, B_{j})$ to each pair of antivirus engines in $A$. Let $\sum D$ denote the sum of all elements in $D$. Because values below the main diagonal of a similarity matrix are redundant, we define the $\triu$($X$, $i$) function to return $X$ where all elements at or below the $i^{th}$ diagonal are replaced with zero. In future references to similarity matrices in this paper it is implicit that redundant information has already been removed, \emph{i.e.} $D$ has been replaced with $\triu$($D$, 1). To measure overall consensus amongst a set of antivirus engines, we use \emph{synchronicity}, defined as \cite{hurier}:

\begin{definition} \label{def:synchronicity}
$\synchronicity(B) = \frac{\sum \triu(D, 1)} {n(n - 1)/2}$
\end{definition}

Synchronicity is equivalent to the average value of the entries above the main diagonal of $D$. We define synchronicity using different notation than \citet{hurier} to to be consistent with  terminology we use later in this paper. When computing the similarity matrix $D$, $\textit{sim}(B_{i}, B_{j})$ can be any pairwise similarity function; we elect to use agreement as this similarity function in all of our experiments. Although \citet{hurier} define synchronicity only for measuring the level of consensus amongst antivirus detections, it can also measure classification consensus by computing a similarity matrix for $C$ instead of $B$.

\begin{wrapfigure}[16]{R}{0.55\textwidth}
\vspace{-12pt}
\centering
\includegraphics[width=0.55\columnwidth,keepaspectratio]{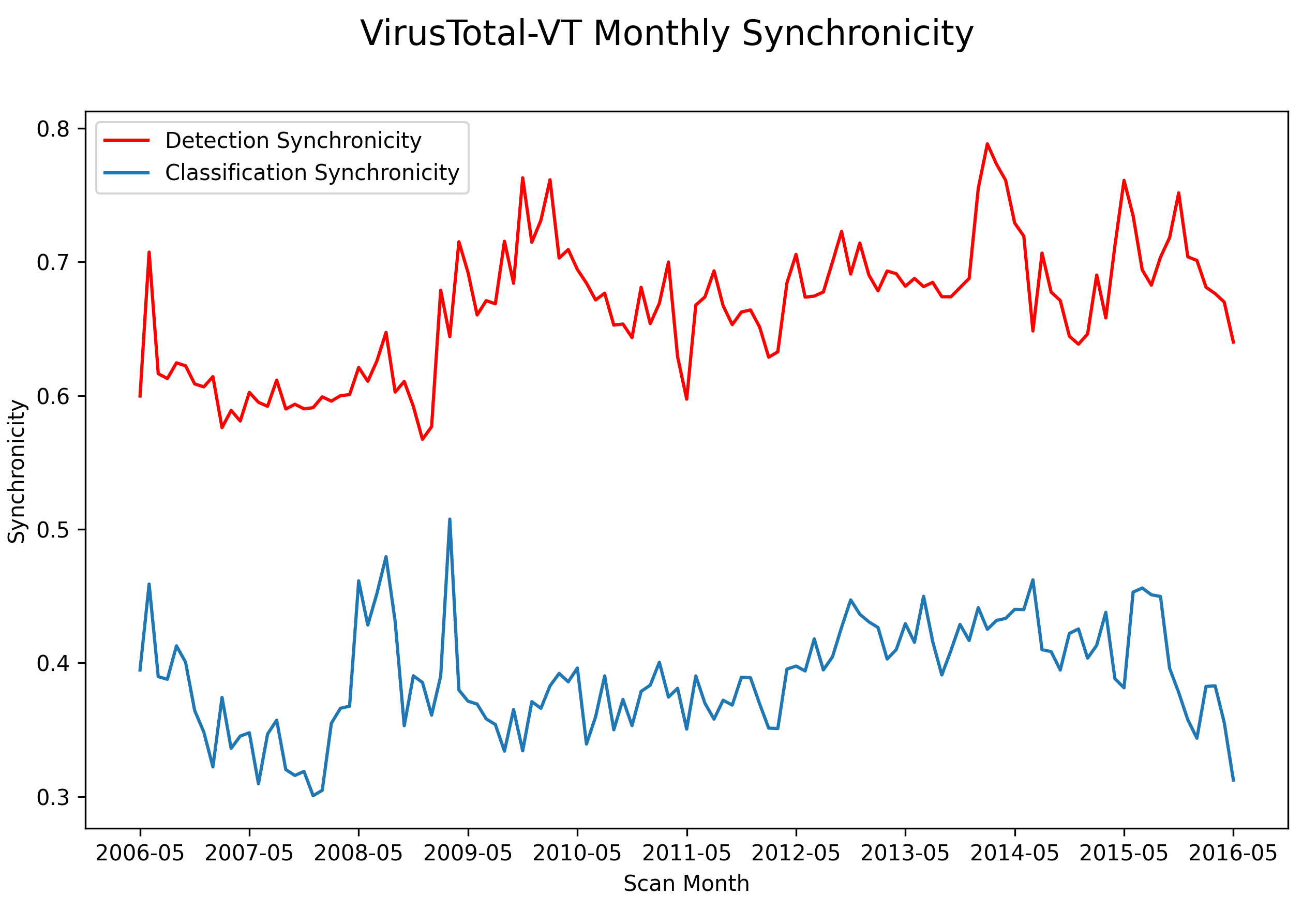}
\caption{Monthly detection and classification synchronicity amongst antivirus engines in VirusShare-VT.}
\label{fig:synchronicity}
\end{wrapfigure}

\subsection{Monthly Synchronicity in VirusShare-VT}
Figure \ref{fig:synchronicity} displays how the synchronicity of the antivirus engines in the VirusShare-VT dataset changes over time. This data was collected by grouping the scans in VirusShare-VT by month and computing detection and classification synchronicity for each group of scans. It is evident that synchronicity amongst antivirus engines varies considerably over short spans of time. Although they have different magnitudes, detection and classification synchronicity seem to be loosely correlated. Again, a possible explanation for this is because classification must follow detection.

At first, we believed that one factor which contributed to the volatility shown in Figure \ref{fig:synchronicity} was engines joining and leaving the VirusTotal platform. As we mentioned in Section \ref{sec:av-consensus}, changes in the set of antivirus engines used by VirusTotal tend to be very gradual. However, we identified three events in which four or more antivirus engines were added or removed in the span of a month. One of these represents the most significant population shift in our dataset by far; the removal of fourteen engines between Jan. and Feb. 2009. This corresponds to an increase in detection synchronicity from 0.577 to 0.679 during this period, though change in classification synchronicity is negligible. The other two events are the additions of four antivirus engines between Aug. and Sep. 2008 and five engines between Aug. and Sep. 2013. However, synchronicity does not change significantly during either of these intervals. It would be difficult for changes in synchronicity to occur due to population changes unless a significant number of engines join or leave. In addition, we observe other significant increases and decreases in synchronicity during which the population of antivirus engines does not change. We conclude that changes in synchronicity amongst antivirus engines are likely caused by a complex assortment of factors, including changes in both the malware ecosystem and antivirus community.

\section{Temporal Rank-1 Decomposition of Similarity Matrices}
\label{sec:r1sm-t}

All current explanations for consensus between antivirus engines can be classified as first-order interactions, \emph{i.e.} a single interaction between a pair of features. In order to test these widely-held assumptions we introduce the Rank-1 Similarity Matrix (R1SM) decomposition. We later describe an extension to R1SM that reveals changes in first-order interactions within time-series data, which we call the Temporal Rank-1 Similarity Matrix Decomposition (R1SM-T). It was necessary to create this decomposition as, to our knowledge, no existing algorithm possesses this capability. 

\subsection{The R1SM Decomposition}
\label{sec:r1sm-decomposition}

Suppose we have a similarity matrix $D$ that represents agreement between each pair of antivirus engines in $A$. The R1SM decomposition exposes first-order interactions between the antivirus engines in the upper triangular of $D$ as the sum of rank-1 outer products with shared, non-negative weights.

\begin{definition} \label{def:R1SM}
$D$ = $\sum_{i=1}^{k} \triu$($\boldsymbol{r}_{i}\boldsymbol{r}_{i}^{\top}, 1$)
\end{definition}

In Definition \ref{def:R1SM}, each vector $\boldsymbol{r}_{1}, \boldsymbol{r}_{2}, ... \boldsymbol{r}_{k}$ has length $n$ and is non-negative. First-order interactions between objects in $D$ manifest in these vectors, which we call the \emph{components} of the decomposition. This behavior occurs due to the nature of the decomposition, in which the outer product of each vector $r_{i}$ and its transpose forms a rank-1 matrix (a matrix containing only first-order interactions by definition). The R1SM decomposition is comparable to the existing CANDECOMP/PARAFAC (CP) decomposition, which also decomposes a tensor into a sum of rank-one outer products \cite{Kolda2009TensorDA}. However, additional restrictions (\emph{e.g.} the decomposition can only be applied to the upper triangular of a square, non-negative matrix and the rank-one outer products have shared weights) distinguish the R1SM decomposition from the CP decomposition. 

\subsection{Solving the R1SM Decomposition}
\label{sec:solving-r1sm}

\algdef{SE}[DOWHILE]{Do}{doWhile}{\algorithmicdo}[1]{\algorithmicwhile\ #1}%

\begin{wrapfigure}[14]{R}{0.5\textwidth}
\vspace{-22pt}
\begin{minipage}{0.5\textwidth}
\begin{algorithm}[H]
\caption{R1SM Greedy Decomposition}
\label{alg:R1SM-greedy}
\begin{algorithmic}[1]
\Require Similarity matrix $D$, early stopping threshold $\delta$

\Function{R1SM-Greedy}{$D$, $\delta$} 
    \State $Y_{1} \leftarrow D$, $i \leftarrow 0$
    \Do
        \State $i \leftarrow i + 1$
        \State Find $r_{i}$ which maximally explains $Y_{i}$ \label{line:ri}
        \State $R_{i} \leftarrow \triu$($\boldsymbol{r}_{i}\boldsymbol{r}_{i}^{\top}$, $1$) \label{line:Ri}
        \State $Y_{i+1} \leftarrow Y_{i} - R_{i}$ \label{line:sub}
    \doWhile $\frac{\sum R_{i}}{\sum D} \geq \delta$ \label{line:loop}
    \State \Return $\boldsymbol{r}_{1}, \boldsymbol{r}_{2}, ... \boldsymbol{r}_{i-1}$
\EndFunction

\end{algorithmic}
\end{algorithm}
\end{minipage}
\end{wrapfigure}

Next, we discuss how the R1SM decomposition is computed. A trivial solution of the R1SM decomposition exists for all similarity matrices in which each component determines a single value in one of the $n(n-1)/2$ elements in the upper triangular. However, this solution does not provide any useful insights about first-order interactions in the decomposed matrix. Recall that one of our research goals is to determine what portion of the correlations can be explained by first-order interactions. In order to obtain this information, we solve for the components of the R1SM decomposition using an iterative, greedy strategy.

Algorithm \ref{alg:R1SM-greedy} approximates the R1SM decomposition of a similarity matrix $D$. At the beginning of the $i^{th}$ iteration of the algorithm, $Y_{i}$ is the residual of $\triu$($D$, 1), representing the portion of the similarity matrix that has not yet contributed to the decomposition. At each step of the decomposition, a component $r_{i}$ is found such that $r_{i}$ maximally explains $Y_{i}$, \emph{i.e.} the maximum value of $\sum R_{i}$ for which $Y_{i} - R_{i}$ is non-negative, where $R_{i} = \triu$($\boldsymbol{r}_{i}\boldsymbol{r}_{i}^{\top}$, $1$) (line \ref{line:ri}). In Section \ref{sec:rism-t-decomposition} we describe our implementation for finding components that maximally explain $Y_{i}$. After solving for $r_{i}$, the updated residual $Y_{i+1}$ is computed by subtracting $R_{i}$ from $Y_{i}$ (line \ref{line:sub}).

Each component of the R1SM decomposition explains a portion of the similarity matrix, given by $\frac{\sum R_{i}}{\sum D}$. Due to the greedy nature of Algorithm \ref{alg:R1SM-greedy}, the percentage of the similarity matrix explained by subsequent components tends to decrease monotonically. Once a component fails to explain a meaningful percentage of the similarity matrix, it is unlikely that any subsequent component will. Once the algorithm reaches this point, we assert that most if not all significant first-order interactions have been captured by the decomposition, and all further information left to be explained is better represented by a more complex model. Therefore, iteration of Algorithm \ref{alg:R1SM-greedy} halts if a component is found for which $\frac{\sum R_{i}}{\sum D}$ is less than $\delta$, which defaults to 0.1\% (line \ref{line:loop}). If a significant portion of $D$ can be decomposed before the early stopping condition is reached, we conclude that most of the interactions between the antivirus engines represented by $D$ are first-order. A complete decomposition of $D$ can be obtained by setting $\delta$ to zero, in which Algorithm \ref{alg:R1SM-greedy} will iterate until $\triu$($Y_{i}$, 1) stores the zero matrix.

\subsection{R1SM Cluster Extraction}
\label{sec:clustering}

Each component $\boldsymbol{r}_i$ of the R1SM composition represents first-order interactions between objects in a similarity matrix. As such, each component can be interpreted as a cluster, where large values in a component indicate a strong first-order relationship between the corresponding objects. Unlike traditional methods for clustering objects in a similarity matrix (e.g. agglomerative hierarchical clustering), which group objects by their overall similarity, the clusters produced by the R1SM decomposition indicate groups with prominent first-order interactions. We stress that clustering is not the primary motivation of the R1SM decomposition, but we explore the idea due to its usefulness.

Because the $\boldsymbol{r}_i$ are not sparse, they may contain small, even spurious values that are not indicative of significant first-order interactions between objects. Thus a parameter $\epsilon$ influences which members of a component are considered ``clustered'' (i.e., a non-trivial first-order correlate). For a component $\boldsymbol{r}_{i}$, the $j$'th object is a member of cluster $i$ iff $\boldsymbol{r}_{ij} \geq \epsilon$. A large $\epsilon$ results in smaller clusters, where all objects within a cluster have strong first-order interactions between each other. Conversely, a small $\epsilon$ yields larger clusters, but objects within a cluster may have weaker first-order interactions. An object may be a member of multiple clusters or none at all, and it is possible for a cluster to contain zero objects. The early stopping term $\delta$ also controls the resulting clustering, as it determines the maximum number of clusters. In Section \ref{sec:clustering-R1SM} we take advantage of the clustering property of the R1SM decomposition to identify groups of antivirus engines that share strong first-order interactions.

\subsection{RISM-T: Applying the R1SM Decomposition to Time-Series Data}
\label{sec:rism-t-decomposition}

One of our primary research goals is to study how first-order interactions amongst antivirus engines have changed over time. In order to do so, we introduce an extension to the R1SM decomposition which can decompose a time-series of similarity matrices $\boldsymbol{D} = [D_{1}, D_{2}, ... D_{T}]$ rather than a single matrix. We call this extension the Temporal Rank-1 Similarity Decomposition (R1SM-T). Again, it is implicit that all similarity matrices in the time-series have had the redundant information at or below their main diagonals replaced with zero.

Algorithm \ref{alg:R1SM-t-greedy} describes the concurrent R1SM decomposition of multiple similarity matrices while sharing information across all matrices as a function of their spatial relationships in time. During the $i^{th}$ iteration, $\boldsymbol{Y}_{i} = [Y_{i,1}, Y_{i,2}, ... Y_{i,T}]$ stores the residual of each similarity matrix in $\boldsymbol{D}$. Like Algorithm \ref{alg:R1SM-greedy}, components $r_{i}$ = [$\boldsymbol{r}_{i,1}, \boldsymbol{r}_{i,2}, ... \boldsymbol{r}_{i,T}$] are found such that they each maximally explain their respective matrices in $\boldsymbol{Y}_{i}$ (lines \ref{line:start-solve} - \ref{line:end-solve}). Our implementation for finding these components is described momentarily. A penalty term discourages any values in $\triu$($\boldsymbol{r}_{i,t}\boldsymbol{r}_{i,t}^{\top}$, $1$) from exceeding their corresponding values in $Y_{i,t}$, but minute errors are still possible. Therefore, for each time $t$, $\triu$($\boldsymbol{r}_{i,t}\boldsymbol{r}_{i,t}^{\top}$, $1$) is corrected using $Y_{i,t}$ and the result is stored in $R_{i,t}$ (line \ref{line:min-t}). $Y_{i+1,t}$, the new residual of $\triu$($D$, 1), is computed by subtracting $R_{i,t}$ from $Y_{i,t}$ (line \ref{line:sub-t}). Like Algorithm \ref{alg:R1SM-greedy}, iteration stops once components are found such that $\frac{\sum R_{i}}{\sum \boldsymbol{D}} < \delta$ (line \ref{line:loop-t}).

Our implementation uses a deep neural network $F(\cdot)$ over positional embeddings to concurrently solve the next component in the R1SM decomposition for each similarity matrix in the time-series. This model design was selected so that non-linear changes in consensus over time can be learned. Furthermore, positional embeddings allow the model to leverage temporal relationships between the target similarity matrices as the primary factor of changes in consensus. $F(\cdot)$ is trained on a batch of input vectors $X = [X_{1}, X_{2}, ..., X_{T}]$, where each vector $X_{t}$ is the positional embedding of timestep $t$ in the time-series. To obtain the positional embedding of $t$, we define $\frac{d}{2}$ distinct frequencies $f_{1}, f_{2}, ... f_{d/2}$, where $d$ is the size of the neural network's input layer and the $j^{th}$ frequency is given by $f_{j} = \frac{t}{10000^{\frac{2j}{d}}}$. $X_{t}$ is constructed by alternately applying the $\sin()$ and $\cos()$ functions to each frequency as shown below \citep{vaswani}.

\theoremstyle{definition}
\begin{definition}
$X_{t} = \left[
           \sin\left(f_{1}\right),
           \cos\left(f_{1}\right),
           \ldots, 
           \sin\left(f_{\frac{d}{2}}\right),
           \cos\left(f_{\frac{d}{2}}\right)
           \right]^\top
$
\end{definition}

\begin{wrapfigure}[26]{R}{0.6\textwidth}
\vspace{-20pt}
\begin{minipage}{0.6\textwidth}
\begin{algorithm}[H]
\caption{R1SM-T Decomposition}
\label{alg:R1SM-t-greedy} 
\begin{algorithmic}[1]
\Require Time-series of similarity matrices $\boldsymbol{D} = [D_{1}, D_{2}, ... D_{T}]$, early stopping threshold $\delta$, and penalty term $\lambda$

\Function{R1SM-T}{$\boldsymbol{D}$, $\delta$, $\lambda$}
    \State $\boldsymbol{Y}_{1} \leftarrow \boldsymbol{D}$, $i \leftarrow 0$
    \Do
        \State $i \leftarrow i + 1$
        \State Initialize network $F(\cdot)$

        \While{F has not converged} \label{line:start-solve}
            \State $\ell \leftarrow 0$ 
            \State $\boldsymbol{r}_{i,1}, \boldsymbol{r}_{i,2}, ... \boldsymbol{r}_{i,T} \leftarrow F(X)$
            \For{$t \leftarrow 1$ to $T$}
                \State $\boldsymbol{U}_{t} \leftarrow \operatorname{min}$($\triu$($\boldsymbol{r}_{i,t}\boldsymbol{r}_{i,t}^{\top}$, $1$) -  $Y_{i,t}$, 0) \label{line:under}
                \State $\boldsymbol{O}_{t} \leftarrow \operatorname{max}$($\boldsymbol{Y}_{i,t}$ - $\triu$($\boldsymbol{r}_{i,t}\boldsymbol{r}_{i,t}^{\top}$, $1$), 0) \label{line:over}
                \State $\ell \leftarrow \ell + \left\|\lambda \boldsymbol{U}_{t} + \boldsymbol{O}_{t}\right\|_{2}$ \label{line:loss}
            \EndFor
            \State Back-propagate $\ell$ and run optimizer step. \label{line:end-solve}
        \EndWhile

        \State $\boldsymbol{r}_{i,1}, \boldsymbol{r}_{i,2}, ... \boldsymbol{r}_{i,T} \leftarrow F(X)$

        \For{$t \leftarrow 1$ to $T$} \label{line:for-3}
            \State $R_{i,t} \leftarrow \operatorname{min}$($\triu$($\boldsymbol{r}_{i,t}\boldsymbol{r}_{i,t}^{\top}$, $1$), $\boldsymbol{Y}_{i,t}$) \label{line:min-t}
            \State $\boldsymbol{Y}_{i+1,t} \leftarrow \boldsymbol{Y}_{i,t} - R_{i,t}$ \label{line:sub-t}
        \EndFor
    \doWhile $\frac{\sum R_{i}}{\sum D} \geq \delta$ \label{line:loop-t}
    \State \Return $\boldsymbol{r}_{1}, \boldsymbol{r}_{2}, ... \boldsymbol{r}_{i-1}$
\EndFunction
\end{algorithmic}
\end{algorithm}
\end{minipage}
\end{wrapfigure}

The use of a single network that predicts a component for each similarity matrix based on positional embeddings permits information sharing across time while simultaneously allowing the model to adjust the results over time. In doing so, the model gains the ability to learn meaningful results during periods in which less data is available, adapting to the rate of change that is present in the data. That is to say, if time is not relevant at all the model can learn to ignore the input embedding $X_t$ entirely. If time is relevant, the embeddings $X_t$ and $X_{t+\Delta}$ have a relationship that can be extracted by a single layer of a neural network \citep{vaswani}, allowing for information sharing over time. This information sharing is important as we observe different rates of change over time, and the amount of samples per month varies by up to three orders of magnitude (as shown in \autoref{fig:datehist}).

During each iteration $i$ a new neural network $F(\cdot)$ optimizes the values in components $r_{i}$ = [$\boldsymbol{r}_{i,1}, \boldsymbol{r}_{i,2}, ... \boldsymbol{r}_{i,T}$] such that they each maximally explain their respective matrices in $\boldsymbol{Y}_{i}$ (lines \ref{line:start-solve} - \ref{line:end-solve}). The loss $\ell$ of $F($X$)$ is computed using two matrices $U_{t}$ and $O_{t}$, which represent element-wise differences between $\triu$($\boldsymbol{r}_{i,t}\boldsymbol{r}_{i,t}^{\top}$, $1$) and $Y_{i,t}$ per timestep. $U_{t}$ stores under-predictions in $\triu$($\boldsymbol{r}_{i,t}\boldsymbol{r}_{i,t}^{\top}$, $1$) (line \ref{line:under}) and $O_{t}$ stores over-predictions in $\triu$($\boldsymbol{r}_{i,t}\boldsymbol{r}_{i,t}^{\top}$, $1$) (line \ref{line:over}). $F(\cdot)$ is strongly discouraged from over-predicting $Y_{i}$ by the $\lambda$ hyper-parameter, which has a value in the range (0, 1], and is set to 0.01 by default. $\lambda$ acts as a scaling factor between $U$ and $O$, causing values in $O$ to contribute more heavily to the loss (line \ref{line:loss}). Due to this term, over-prediction of the values in the components is rare. Once the batch loss has been computed, the model performs back-propagation and the optimizer step (line \ref{line:end-solve}). Training continues until the model converges, at which point $\boldsymbol{r}_{i}$ holds an optimal solution. Algorithm \ref{alg:R1SM-t-greedy} can solve the R1SM decomposition of a single similarity matrix by defining it as a time-series with only one timestep. 
Our implementation of the neural network $F(\cdot)$ uses ten hidden layers with five residual connections. The default hidden layer size is 1,024 neurons, and the network includes multiple bottleneck layers whose sizes are a function of the input and output layer sizes.
\begin{wrapfigure}[18]{R}{0.5\textwidth}
\vspace{16pt}
    \centering
    \includegraphics[width=0.47\columnwidth,keepaspectratio]{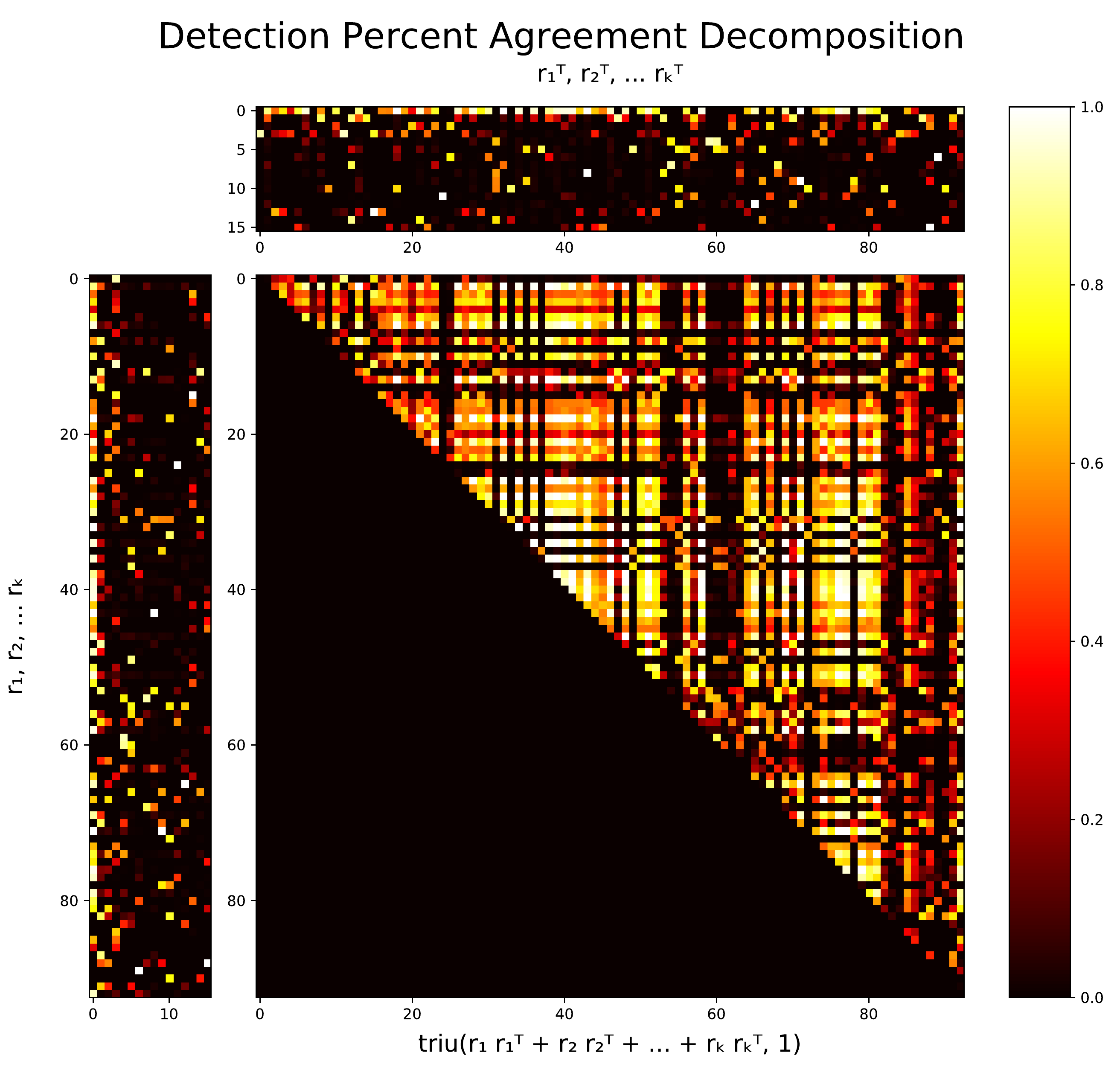}
    \caption{R1SM decomposition of the detection agreement similarity matrix for VirusShare-VT. The leftmost subplot contains the sixteen components and the topmost subplot contains their transposes. The large central subplot contains the portion of the similarity matrix explained by the components.}
    \label{fig:components_detect}
\end{wrapfigure}
The $\operatorname{exp()}$ function is applied to all weights in the output layer of $F(\cdot)$, constraining the predicted components to be non-negative as required by the definition of the R1SM decomposition. An important design factor is the use of a very small learning rate, which allows precise adjustments to the values in the component during the learning process. By default, R1SM-T uses a learning rate of 1e-7. 
We note that in extended tests a wide array of layer depths, residual and/or simple feed-forward connections, and numbers of neurons per layer, produced qualitatively and quantitatively the same results, as the networks are learning to predict population level statistics without explicit features about the populations, forcing the network to learn consistent population behaviors.

\vspace*{-10pt}
\leavevmode\section{First-Order Interactions in Antivirus Scan Data}
\label{sec:av}

Now that we have introduced the R1SM decomposition and R1SM-T, we use them to study first-order interactions amongst the antivirus engines in the VirusTotal-VT dataset. First, we investigate the validity of the industry assumption that consensus between antivirus engines is caused by first-order interactions, such as sharing of threat intelligence and copying from leading vendors. We identify clusters of antivirus engines with strong first-order interactions. Finally, we research how first-order interactions between antivirus engines have changed over a decade.

\subsection{Applying R1SM to Antivirus Similarity Matrices}
\label{sec:clustering-R1SM}
\begin{wrapfigure}[16]{R}{0.5\textwidth}
\vspace{2pt}
    \centering
    \includegraphics[width=0.45\columnwidth,keepaspectratio]{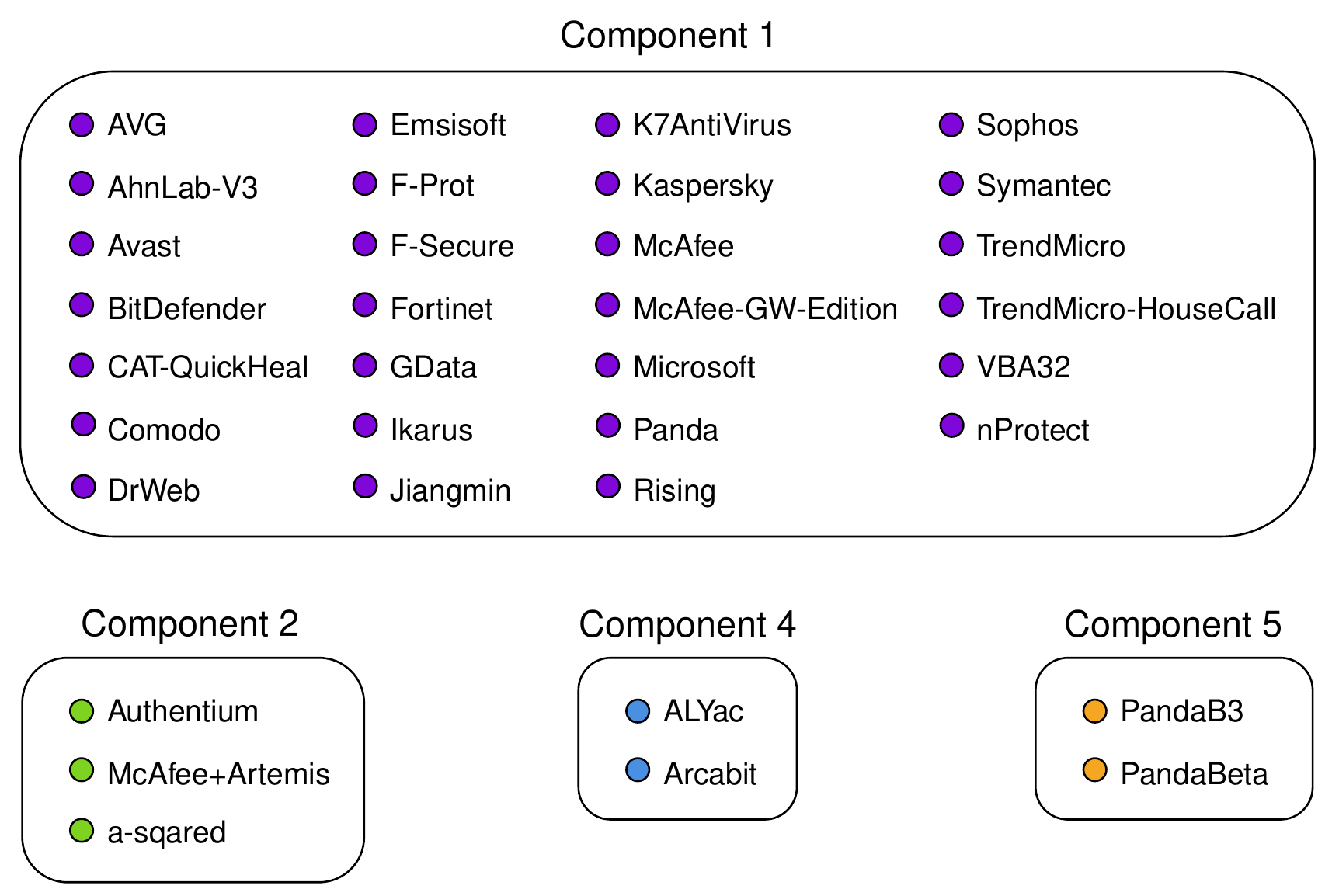}
    \caption{Clusters extracted from the R1SM decomposition of the detection percent agreement matrix ($\delta=0.1\%, \epsilon=0.85$).}
    \label{fig:detection-clusters}
\end{wrapfigure}

Figure \ref{fig:components_detect} displays the R1SM decomposition of

        \noindent the similarity matrix shown in Figure \ref{fig:heatmap_detect}, which measures pairwise detection agreement amongst the antivirus engines in the VirusShare-VT dataset. This decomposition was obtained by applying Algorithm \ref{alg:R1SM-t-greedy} to the similarity matrix, represented as a time-series with a single timestep. Using an early stopping threshold of $\delta = 0.1\%$, the decomposition yielded $k = 16$ components which explain 60.596\% of the matrix. That approximately 40\% of the matrix went unexplained implies that significant amounts of the consensus between antivirus engines cannot be explained by first-order interactions alone, which runs 

\clearpage 
\noindent counter to current belief. Determining the 
\begin{wrapfigure}[18]{R}{0.5\textwidth}
    \vspace{-28pt}
    \centering
    \includegraphics[width=0.5\columnwidth,keepaspectratio]{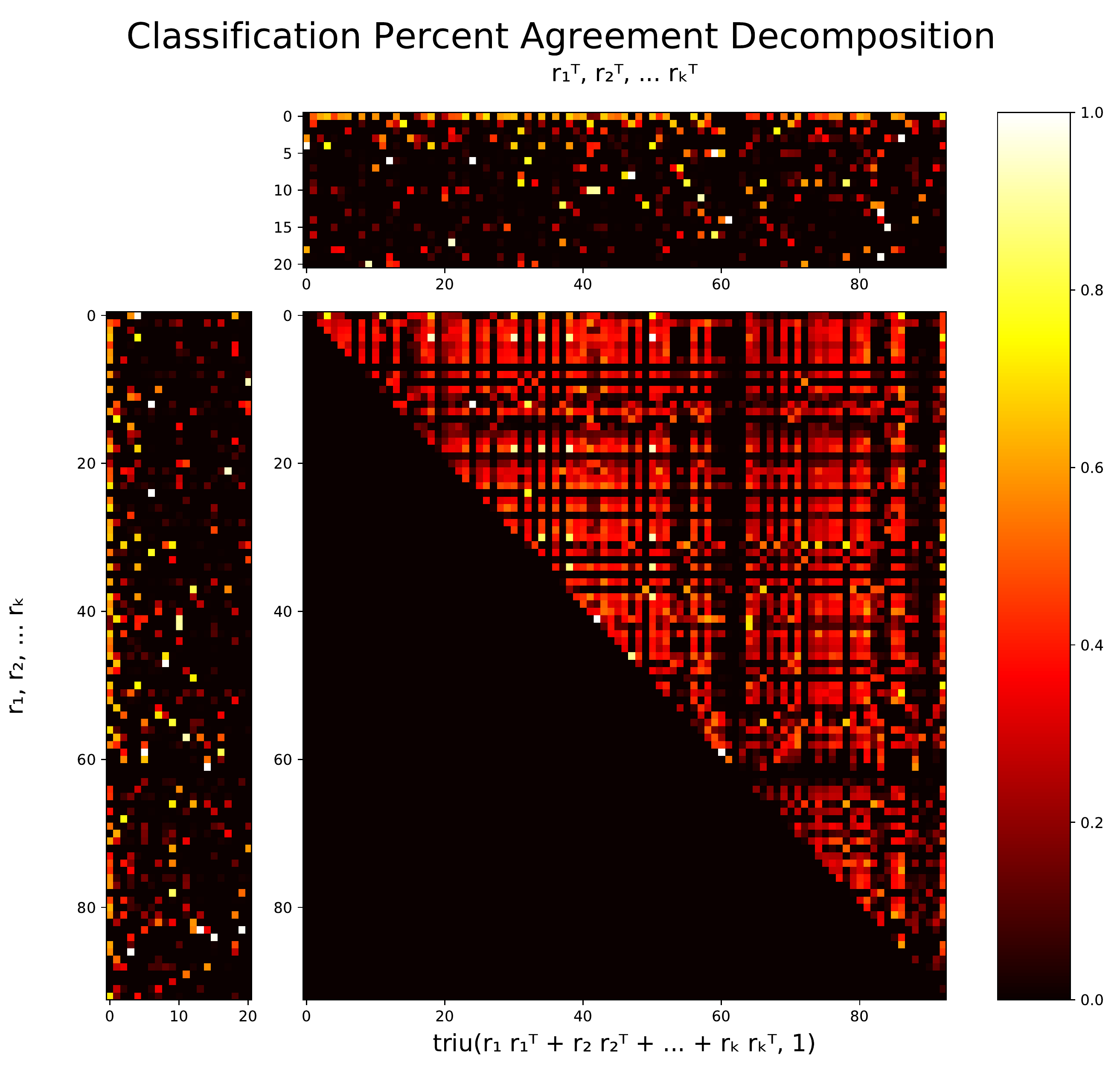}
    \caption{R1SM decomposition of the antivirus classification agreement similarity matrix.}
    \label{fig:components_family}
\end{wrapfigure}

\noindent nature of such interactions is a subject for future research.

Figure \ref{fig:detection-clusters} displays clusters extracted from the R1SM decomposition in Figure \ref{fig:components_detect} using $\epsilon$ = 0.85. Components with less than two antivirus engines exceeding $\epsilon$ are not shown. The clustering illustrates a common trait of the R1SM decomposition, namely that the first component tends to subsume a large quantity of the similarity matrix, resulting in a large cluster for the first component. The cluster extracted from the first component indicates that a significant number of first-order interactions exist between a large group of antivirus engines. Inspection of the clustering shows pairs of antivirus engines with a shared vendor, such as TrendMicro and TrendMicro-Housecall as well as PandaB3 and PandaBeta. Other antivirus 

\begin{wrapfigure}{R}{0.5\textwidth}
    \vspace{-16pt}
    \centering
    \includegraphics[width=0.5\columnwidth,keepaspectratio]{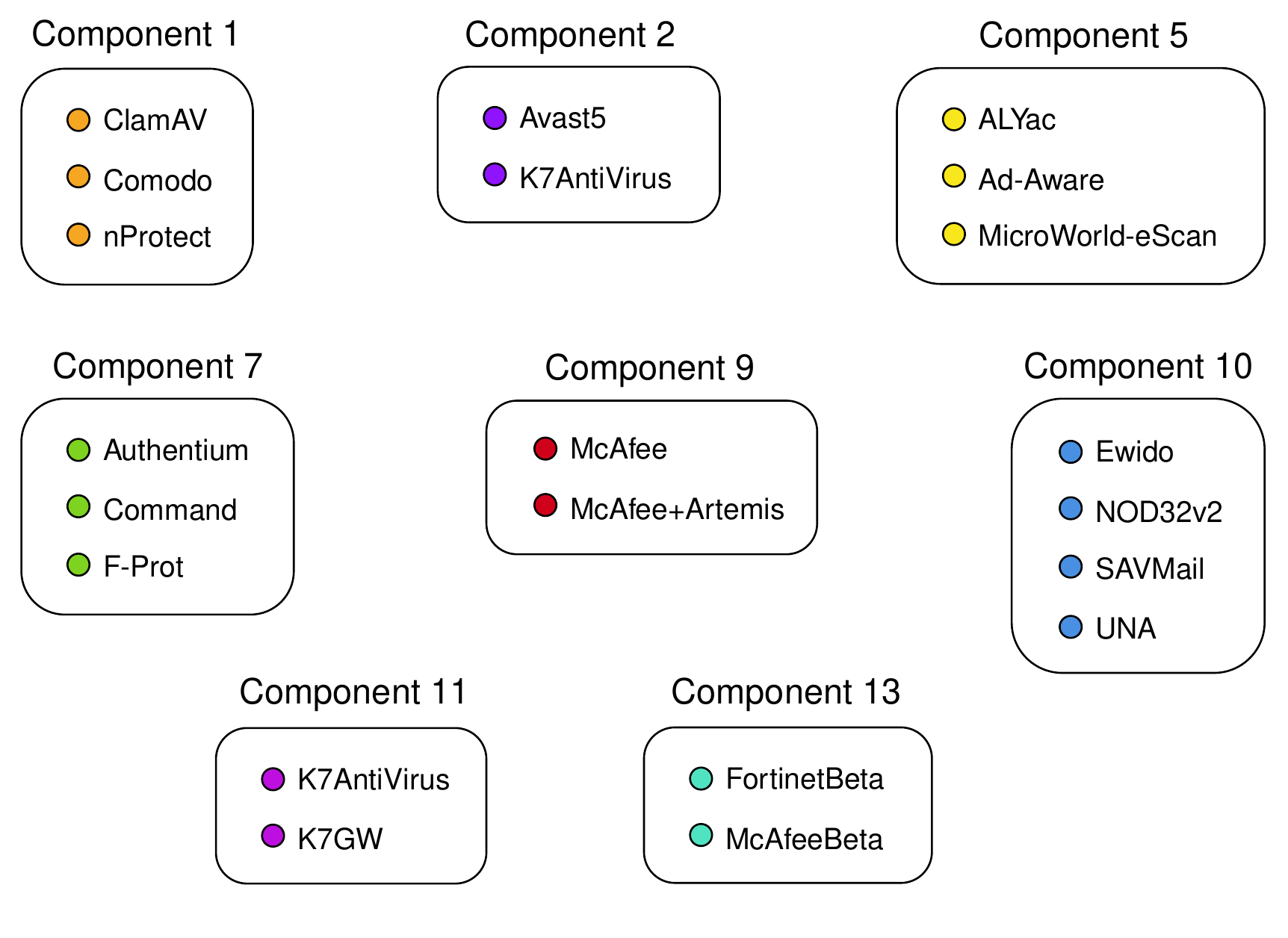}
    \caption{Clusters extracted from the R1SM decomposition of the classification percent agreement matrix ($\delta=0.1\%, \epsilon=0.7$).}
    \label{fig:classification-clusters}
\end{wrapfigure}

\noindent engines in the clusters have been previously reported to have similarities, such as BitDefender, Emsisoft, and GData; McAfee, McAfee-GW-Edition, and Microsoft; and Avast, AVG, and Fortinet \citep{zhu, avclass-common}. Further investigation is needed to identify the causes of the first-order interactions between the remaining antivirus engines.  

Figure \ref{fig:components_family} shows the R1SM decomposition of the similarity matrix shown in Figure \ref{fig:heatmap_family}, which contains pairwise classification agreement scores for the antivirus engines in VirusShare-VT. This decomposition has $k = 21$ components which explain 58.394\% of the matrix. As with the prior decomposition, a significant portion of the similarity matrix cannot be explained using first-order interactions alone, and further work is necessary to identify and model the complex relationships between this set of antivirus engines. Comparing the central subplots of figures \ref{fig:components_detect} and \ref{fig:components_family} shows that both decompositions are structurally alike, indicating that many of the same first-order interactions exist between the antivirus engines whether measuring detection or classification agreement. We revisit this observation in Section \ref{r1sm-t-antivirus}, where we show that the time-series for the two similarity metrics also have R1SM-T decompositions with notable similarities.

Figure \ref{fig:classification-clusters} shows the clusters extracted from the classification percent agreement R1SM decomposition in Figure \ref{fig:components_family} using $\epsilon$ = 0.7. Again, components with less than two antivirus engines exceeding $\epsilon$ are not displayed. Shared vendor relationships between Authentium and Command, McAfee and McAfee+Artemis, and K7AntiVirus and K7GW are identified by the clusters for components 7, 9, and 11 respectively. \citet{zhu} identify similarities between ClamAV and Comodo (component 1) as well as Ad-Aware and MicroWorld-eScan (component 5). \citet{avclass-common} also report that the Ad-Aware and MicroWorld-eScan engines frequently have identical labels. No prior work has identified similarities between any antivirus engines developed by Fortinet and McAfee, but in 2019 the two vendors released a joint endpoint security solution \citep{fortinet-mcafee}. A partnership between Fortinet and McAfee likely accounts for the first-order interactions between their two beta engines in component 13. We have not found any publicly known connections between the remaining clustered antivirus engines.

\subsection{Applying R1SM-T to Time-Series of Antivirus Similarity Matrices}
\label{r1sm-t-antivirus}

Next, we investigate the changes in first-order interactions between antivirus engines in the VirusShare-VT dataset over the course of a decade. To do this, we separated VirusShare-VT into groups of antivirus scans by month, and computed detection and classification agreement similarity matrices for each group. The similarity matrices were then arranged into two time-series representing monthly change in classification and detection agreement respectively. Finally, we applied R1SM-T to both time-series.

\begin{wrapfigure}{R}{0.5\textwidth}
\vspace*{-12pt}
\centering
\includegraphics[width=0.5\columnwidth,keepaspectratio]{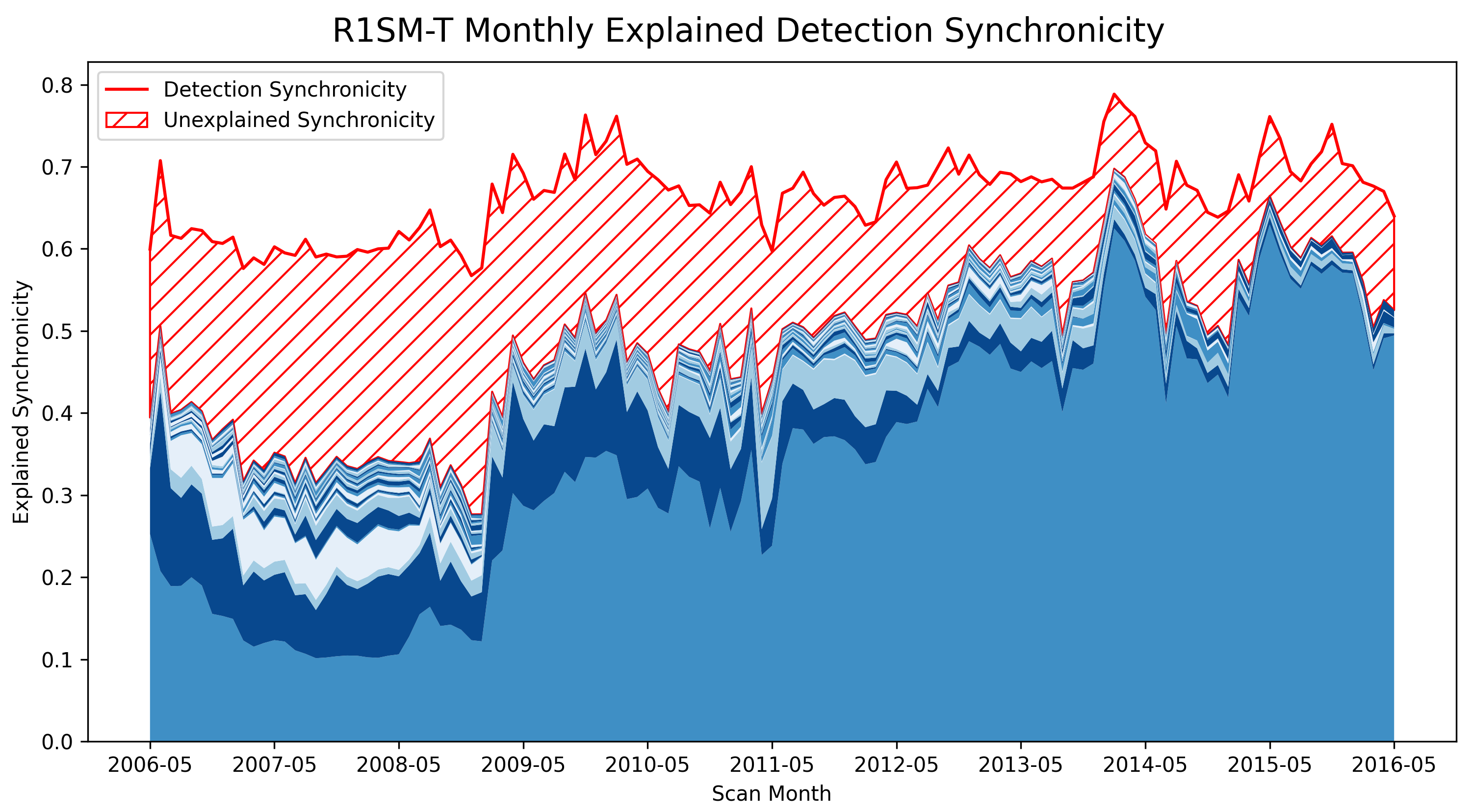}
\includegraphics[width=0.5\columnwidth,keepaspectratio]{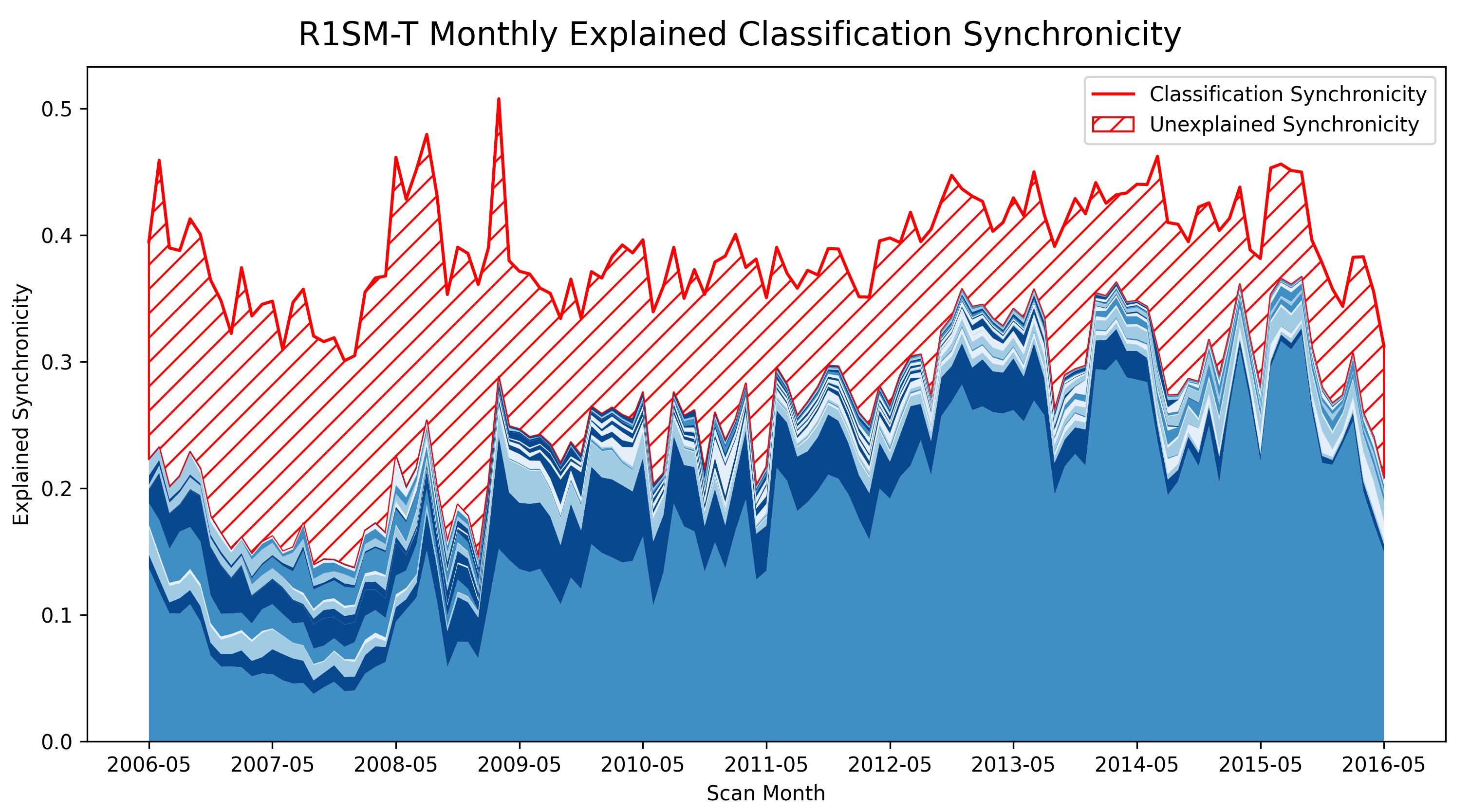}
\caption{Monthly detection and classification synchronicity explained by R1SM-T. Synchronicity contributed by each component is shown in a shade of blue, with component 1 at the bottom. The total monthly synchronicity is indicated by the topmost red line. The red shaded region indicates how much synchronicity is not explained by first-order interactions contained within R1SM-T components.
}
\label{fig:detect-expl-synchronicity}
\end{wrapfigure}

The R1SM-T models for the detection and classification agreement time-series converged after 5,200,000 and 5,440,000 training iterations respectively. They each identified $k = 26$ sets of components using the early stopping value $\delta = 0.1\%$. The R1SM-T decomposition for the detection percent agreement time-series explains an average of 73.709\% of the matrices and the decomposition for the classification percent agreement time-series explains an average of 67.196\% of the matrices. Interestingly, the percent explained by the R1SM-T decomposition varies monthly, as shown in Figure \ref{fig:detect-expl-synchronicity}. In this figure, the upper red line of each plot indicates monthly changes in synchronicity, originally shown in Figure \ref{fig:synchronicity}. Each region shaded in blue represents how much a component of the decomposition contributes to the monthly synchronicity, given by $\frac{\sum R_{i,t}}{n(n-1)/2}$. Synchronicity that cannot be explained by first-order interactions captured in the decomposition are represented by the area shaded in red. In both plots, the proportion of synchronicity explained by first-order interactions slowly increases. Although the cause of this trend is unknown, a possible explanation is an increase in sharing of threat intelligence throughout the industry over time. In both plots, the first component steadily becomes the dominant contributor to the explained synchronicity over time. Before 2009, the other components supplied approximately half of the explained synchronicity, but they became negligible by 2014. This seems to indicate that sharing of threat intelligence used to be limited to disparate groups of antivirus engines, but over time information sharing has become ubiquitous. This also correlates with usage of VirusTotal itself within industry, as it provides extensive threat intelligence tooling and a community-based platform for sharing information about malware samples.

\begin{wrapfigure}{R}{0.5\textwidth}
    \centering
    \adjustbox{width=8.75cm}{%
    \input{detection-components-ts-anno}
    }
    \caption{The first components for the time-series of similarity matrices measuring monthly antivirus detection percent agreement in VirusShare-VT. Each column is the component for a particular month, starting in May 2006 and ending in May 2016.}
    \label{fig:components_detect_ts}

    \vspace*{16pt}
    \adjustbox{width=8.75cm}{%
    \input{classification-components-ts-anno}
    }
    \caption{The first components for the time-series measuring antivirus classification percent agreement in VirusShare-VT.}
    \label{fig:components_family_ts}

\end{wrapfigure}

Next, we investigate the first R1SM-T component of both time-series due to its intriguing behavior in Figure \ref{fig:detect-expl-synchronicity}. In doing so, we observe how the behaviors of individual antivirus engines as well as overall trends in the antivirus community change over time. Figures \ref{fig:components_detect_ts} and \ref{fig:components_family_ts} display the first component of the R1SM decomposition for each of the 121 similarity matrices in the two time-series. Each column represents the component for a particular month, and each row indicates how the contributions of a specific antivirus engine to the first component have changed over time. 

The overall magnitude of the components within Figure
\ref{fig:components_family_ts} is lower than their counterparts in Figure \ref{fig:components_detect_ts}, and month-to-month component values have more variability. However, the similarity in structure between the two decompositions is striking. As with our earlier findings for the two R1SM decompositions, a possible explanation for this structural similarity is that classification depends upon detection. These results could also indicate that the same types of first-order interactions tend to exist between antivirus engines regardless of whether detection or classification agreement is measured. Next, we discuss notable types of features visible in the decomposition that indicate changes in first-order interactions between antivirus engines.

\textbf{Trivial Patterns:} Insights into alterations in antivirus behavior can be observed when corresponding values in the decomposition change radically within a short time period. Both decompositions clearly display the months during which antivirus engines were added to the VirusTotal platform, such as Alyac in Nov. 2014 (row 0) \citep{vt-alyac}. The Jun. 2015 retirement of the Norman antivirus engine from VirusTotal is also visible in both decompositions (row 56) \citep{vt-norman}.

\textbf{Abnormal Structural Changes:} Vertical "bands" in the R1SM-T decompositions indicate periods of change within the entire antivirus community that have never been previously noted or identified, to the best of our knowledge. A band evident in both figures \ref{fig:components_detect_ts} and \ref{fig:components_family_ts} takes place during Apr. and May 2011 (columns 59 and 60), in which values for a number of antivirus engines, including Avast (row 13), Emsisoft (row 30), F-Prot (row 32), GData (row 38), Ikarus (row 39), Rising (row 65), Sophos (row 69), TheHacker (row 74), VIPRE (row 80) drop sharply. A second band beginning in Jul. 2014, which lasts until Feb. 2015 in Figure \ref{fig:components_detect_ts} and until May 2015 in Figure \ref{fig:components_family_ts}, indicates a turbulent period where the relationships between antivirus engines were in flux. The components in Figure \ref{fig:components_family_ts} immediately following this band change drastically, with many antivirus engines gaining an increased share of the component in comparison to the prior months. To understand the cause of these community-wide disturbances in correlation requires further research, but should immediately impact how industry design their label aggregation pipelines. We would recommend any training data labeled during these time periods be regarded as potentially suspect, and such data should undergo further analysis to confirm label quality.

\textbf{Sudden Variations:}
Individual changes to an antivirus within a short period of time also indicate notable events. In Figure \ref{fig:components_detect_ts} a large gap occurs for K7Antivirus (row 41) from Feb. 2010 to Jul. 2010, which corresponds with the release of K7 TotalSecurity version 10.0 on Feb. 23, 2010 \citep{k7-v10}. Aegislab (row 4) fluctuates significantly for unknown reasons, dropping from 0.575 when it was first introduced to VirusTotal in Feb. 2014 \citep{vt-aegislab} to 0.146 and rising back to a peak of 0.716 in Aug. 2014. Aegislab's contributions to the first component are nearly identical to those of Alibaba (row 7) throughout all of 2015, possibly indicating a common information source.

\textbf{Differences Between Decompositions:} Since the first components of both R1SM-T decompositions are structurally very similar, differences between the two may indicate first-order correlations caused by factors related to either benign/malicious detection or family classification alone. These factors could include increased or reduced use of heuristic antivirus signatures or changes in malware family naming conventions. External events, such as the emergence of new malware families, could also explain these discrepancies.

The R1SM-T decompositions in Figures \ref{fig:components_detect_ts} and \ref{fig:components_family_ts} reveal that correlations between antivirus engines can change significantly within a short time period. Furthermore, they illustrate periods of industry-wide change that have never been previously identified. Although we explain many of the features in the decompositions, the factors that cause consensus between antivirus engines to change are still largely unknown, and identifying the sources that cause periods of population-wide volatility is especially important.

\section{Discussion}
\label{sec:conclusion}

We lack complete understanding of the factors that cause correlations between antivirus engines; first-order interactions alone are not sufficient for modeling the complex interconnections between antivirus engines. In studying how consensus amongst antivirus engines change over time, we found that the relationships between antivirus engines are even more intricate and volatile than previously thought. The overall level of consensus amongst antivirus engines can change quickly in short periods of time for reasons which are still not fully understood. Using R1SM-T we found that first-order interactions have become increasingly responsible for consensus between antivirus engines over time, although they are still insufficient for modeling some of the sources of antivirus correlations. Furthermore, we found that first-order interactions now seem to be nearly ubiquitous across the entire antivirus industry, whereas disparate segments of the industry previously existed where first-order interactions could not be identified. Finally, we showed that components of R1SM-T could be utilized to identify individual and population-wide changes in antivirus behavior.

Current understanding of antivirus dynamics is clearly insufficient and more research about the causes of antivirus correlation is needed. It is difficult to trust antivirus results when the factors that cause them to be correlated are still poorly understood. On account of this, and because relationships between antivirus engines can change significantly in a short period of time, existing methods for aggregating antivirus signatures for the purposes of malware detection and classification are flawed. Future aggregation approaches should consider weighted ensembles where the weights of the voting members are also a function of time. We also hope that elements of this work, such as the ability to quantify first-order relationships and assess changes in these relationships over time, may themselves contribute towards improvements in antivirus aggregation. 
\clearpage

\begin{appendices}

\begin{table}[!tbh]
\section{}
\vspace*{0.1cm}
\centering

\adjustbox{width=0.9\columnwidth}{%
\begin{tabular}{@{}rlrrlr@{}}
\toprule
\multicolumn{1}{c}{Index} & \multicolumn{1}{c}{Antivirus Engine} & \multicolumn{1}{c}{Scan Count} & \multicolumn{1}{c}{Index} & \multicolumn{1}{c}{Antivirus Engine} & \multicolumn{1}{c}{Scan Count} \\ \midrule
0                         & ALYac                                & 4,679,821                      & 47                        & McAfee+Artemis                       & 995,699                        \\
1                         & AVG                                  & 24,982,795                     & 48                        & McAfee-GW-Edition                    & 24,607,621                     \\
2                         & AVware                               & 5,664,526                      & 49                        & McAfeeBeta                           & 95,784                         \\
3                         & Ad-Aware                             & 12,649,803                     & 50                        & MicroWorld-eScan                     & 19,382,987                     \\
4                         & AegisLab                             & 10,636,692                     & 51                        & Microsoft                            & 24,984,940                     \\
5                         & Agnitum                              & 19,009,698                     & 52                        & NANO-Antivirus                       & 18,763,016                     \\
6                         & AhnLab-V3                            & 23,792,676                     & 53                        & NOD32                                & 4,738,012                      \\
7                         & Alibaba                              & 4,657,472                      & 54                        & NOD32Beta                            & 343,198                        \\
8                         & AntiVir                              & 19,225,387                     & 55                        & NOD32v2                              & 245,233                        \\
9                         & Antivir7                             & 8,710                          & 56                        & Norman                               & 21,187,821                     \\
10                        & Antiy-AVL                            & 24,559,174                     & 57                        & PCTools                              & 11,426,342                     \\
11                        & Arcabit                              & 3,770,802                      & 58                        & Panda                                & 24,713,903                     \\
12                        & Authentium                           & 1,778,326                      & 59                        & PandaB3                              & 2,695                          \\
13                        & Avast                                & 24,945,279                     & 60                        & PandaBeta                            & 288,403                        \\
14                        & Avast5                               & 2,405,851                      & 61                        & PandaBeta2                           & 3,371                          \\
15                        & Avira                                & 5,432,038                      & 62                        & Prevx                                & 3,826,154                      \\
16                        & Baidu                                & 752,127                        & 63                        & Prevx1                               & 438,718                        \\
17                        & Baidu-International                  & 13,770,014                     & 64                        & Qihoo-360                            & 11,703,897                     \\
18                        & BitDefender                          & 25,037,371                     & 65                        & Rising                               & 24,233,086                     \\
19                        & Bkav                                 & 13,155,628                     & 66                        & SAVMail                              & 149,834                        \\
20                        & ByteHero                             & 20,476,926                     & 67                        & SUPERAntiSpyware                     & 23,567,247                     \\
21                        & CAT-QuickHeal                        & 25,048,885                     & 68                        & SecureWeb-Gateway                    & 101,352                        \\
22                        & CMC                                  & 11,819,328                     & 69                        & Sophos                               & 24,540,725                     \\
23                        & ClamAV                               & 25,007,198                     & 70                        & Sunbelt                              & 1,741,218                      \\
24                        & Command                              & 250,234                        & 71                        & Symantec                             & 24,732,139                     \\
25                        & Commtouch                            & 17,141,359                     & 72                        & T3                                   & 18,561                         \\
26                        & Comodo                               & 24,666,456                     & 73                        & Tencent                              & 8,606,167                      \\
27                        & Cyren                                & 5,885,738                      & 74                        & TheHacker                            & 25,083,441                     \\
28                        & DrWeb                                & 24,599,303                     & 75                        & TotalDefense                         & 20,294,251                     \\
29                        & ESET-NOD32                           & 19,964,248                     & 76                        & TrendMicro                           & 24,752,548                     \\
30                        & Emsisoft                             & 22,526,376                     & 77                        & TrendMicro-HouseCall                 & 23,560,584                     \\
31                        & Ewido                                & 292,281                        & 78                        & UNA                                  & 44,282                         \\
32                        & F-Prot                               & 25,036,409                     & 79                        & VBA32                                & 24,870,783                     \\
33                        & F-Prot4                              & 26,895                         & 80                        & VIPRE                                & 23,181,524                     \\
34                        & F-Secure                             & 24,060,788                     & 81                        & ViRobot                              & 24,843,379                     \\
35                        & FileAdvisor                          & 129,111                        & 82                        & VirusBuster                          & 5,365,970                      \\
36                        & Fortinet                             & 25,045,656                     & 83                        & Webwasher-Gateway                    & 202,898                        \\
37                        & FortinetBeta                         & 89,667                         & 84                        & Yandex                               & 564,996                        \\
38                        & GData                                & 24,831,415                     & 85                        & Zillya                               & 8,444,657                      \\
39                        & Ikarus                               & 25,087,466                     & 86                        & Zoner                                & 7,351,648                      \\
40                        & Jiangmin                             & 24,011,547                     & 87                        & a-squared                            & 1,135,672                      \\
41                        & K7AntiVirus                          & 24,510,017                     & 88                        & eSafe                                & 9,936,896                      \\
42                        & K7GW                                 & 16,303,096                     & 89                        & eScan                                & 37,174                         \\
43                        & Kaspersky                            & 24,721,716                     & 90                        & eTrust-InoculateIT                   & 35,112                         \\
44                        & Kingsoft                             & 17,919,699                     & 91                        & eTrust-Vet                           & 4,731,780                      \\
45                        & Malwarebytes                         & 18,710,973                     & 92                        & nProtect                             & 24,252,689                     \\
46                        & McAfee                               & 24,949,683                     &                           &                                      &                                \\ \bottomrule
\end{tabular}
}
\caption{Antivirus engines present in at least 1,000 scan reports in VirusTotal-VT. The Index column displays which row and/or column in Figures \ref{fig:heatmap_detect}, \ref{fig:heatmap_family}, \ref{fig:components_detect}, \ref{fig:components_family}, \ref{fig:components_detect_ts}, and \ref{fig:components_family_ts} each engine corresponds to.}
\label{tab:av-index}
\end{table}

\end{appendices}

\bibliography{citations}

\end{document}